\documentclass[subeqn,12pt,a4paper]{article} 

\usepackage{amssymb,amsmath,amsfonts,amsthm, amscd, mathrsfs,helvet, mathtools}
\usepackage{bm, stmaryrd}
\usepackage{graphicx,verbatim}
\usepackage{psfrag}
\usepackage[all]{xy}

\theoremstyle{plain}
\newtheorem{theorem}{Theorem}[section]
\newtheorem*{theorem*}{Theorem}
\newtheorem{proposition}[theorem]{Proposition}
\newtheorem*{proposition*}{Proposition}

\newtheorem{lemma}[theorem]{Lemma}
\newtheorem*{lemma*}{Lemma}

\newcommand{\tensor}[1]{{\bf \underline{#1}}}

\DeclareMathOperator{\res}{res}

\def\Loop{\mathcal{L}}
\def\g{\mathfrak{g}} \def\G{\mathfrak{G}}  \def\h{\mathfrak{h}}  \def\Gd{\mathfrak{G}^{\ast}}
\def\hG{\hat{\mathfrak{G}}}
\def\hGd{\hat{\mathfrak{G}}^{\ast}}
\def\gcirc{\mathring{\g}}
\def\gO{\g^{\mbox{\scriptsize $\Omega$}}}

\newcommand{\bb}[1]{\llbracket #1 \rrbracket}

\def\A{\mathcal{A}}
\def\C{\mathcal{C}}
\def\O{\mathcal{O}}
\def\P{\mathcal{P}}
\def\L{\mathfrak{L}}
\def\M{\mathfrak{M}}
\def\N{\mathfrak{N}}

\def\X{\mathfrak{X}}
\def\Y{\mathfrak{Y}}

\def\1{\tensor{1}}
\def\2{\tensor{2}}
\def\3{\tensor{3}}

\numberwithin{equation}{section}

\begin{document}

\begin{titlepage}
\begin{flushright}
IPhT-t10/026\\
\end{flushright}
\begin{centering}
\vspace{10mm} %
{\Large {\bf The classical $R$-matrix of AdS/CFT\\ \vspace{2mm} and its Lie dialgebra structure}}\\

\vspace{12mm}
{\large Beno\^{\i}t Vicedo}\\
\vspace{8mm}

{\it Laboratoire de Physique Th\'eorique, \'Ecole Normale Sup\'erieure\\
24 rue Lhomond, 75231 Paris CEDEX-5, France}\\
\vspace{4mm}
and\\
\vspace{4mm}
{\it Institut de Physique Th\'eorique, C.E.A. - Saclay,\\
F-91191 Gif-sur-Yvette, France}\\
\vspace{2mm}
\small{\tt benoit.vicedo@cea.fr}

\vspace{10mm} 

\begin{abstract}
\centering \hspace{-7mm}
\begin{minipage}{0.8\textwidth}
\vspace{5mm}
The classical integrable structure of $\mathbb{Z}_4$-graded supercoset $\sigma$-models, arising in the AdS/CFT correspondence, is formulated within the $R$-matrix approach. The central object in this construction is the standard $R$-matrix of the $\mathbb{Z}_4$-twisted loop algebra. However, in order to correctly describe the Lax matrix within this formalism, the standard inner product on this twisted loop algebra requires a further twist induced by the Zhukovsky map, which also plays a key role in the AdS/CFT correspondence. The non-ultralocality of the $\sigma$-model can be understood as stemming from this latter twist since it leads to a non skew-symmetric $R$-matrix.
\end{minipage}
\end{abstract}

\end{centering}

\end{titlepage}

\input{epsf}

\section{Introduction and summary} \label{sect: 0}

Since the early days of the ongoing study of the AdS$_5$/CFT$_4$ conjecture \cite{AdSCFT} in the planar sector, footprints of integrability were discovered both in gauge theory \cite{Gauge Integrability 1, Gauge Integrability 2, Gauge Yangian} where the 't Hooft coupling $\lambda \ll 1$ is very small as well as in string theory \cite{String Integrability} where $\lambda \gg 1$, hinting at the existence of a deep underlying integrable structure at all intermediate values $\lambda \sim 1$. In light of these encouraging discoveries, the generally favoured approach thus far has been a top-down one, whereby quantum integrability was assumed and its implications could be subsequently tested against perturbative data from both sides of the correspondence.

Indeed, in the infinite volume limit $L \gg 1$ where scattering makes sense, the notion of quantum integrability in $(1+1)$-dimensions essentially reduces to the factorisability of the $\mathcal{S}$-matrix. Thus regardless of the details of the hypothetical integrable structure at $\lambda \sim 1$, in the limit $L \rightarrow \infty$ the central object becomes the $2 \to 2$ $\mathcal{S}$-matrix. In the fundamental representation, the latter could be determined from its invariance under the maximal central extension $\mathfrak{h} \coloneqq \mathfrak{psu}(2|2) \ltimes \mathbb{R}^3$ of the residual symmetry algebra \cite{Smatrix} and such that it satisfies the requirements of unitarity, crossing symmetry and the quantum Yang-Baxter equation \cite{Crossing}. When expressed in terms of $\mathcal{R} = P \mathcal{S}$, where $P$ is a permutation, these properties suggest the existence of a subjacent Yangian based on $\mathfrak{h}$ \cite{Beisert:2007ds}. Yet no universal $\mathcal{R}$-matrix for the non-semisimple algebra $\mathfrak{h}$ has yet been identified. However, the classical limit $\lambda \gg 1$ of this hypothetical universal $\mathcal{R}$-matrix, namely a universal classical $r$-matrix with its associated Lie bialgebra structure, was conjectured in \cite{Beisert-Spill}.

On the CFT side of the story, the 1-loop dilatation operator of planar $\mathcal{N} = 4$ SYM theory was identified with the Hamiltonian of an integrable spin-chain \cite{Gauge Integrability 1}, a property which seems to persist also at higher loops \cite{Gauge Integrability 2}. 
Subsequent investigations of the integrable structure of planar $\mathcal{N} = 4$ SYM theory revealed that the dilatation operator admits, at the first few orders of perturbation theory, a Yangian symmetry based on the full $\mathfrak{psu}(2,2|4)$ algebra \cite{Gauge Yangian}. This symmetry, however, is exact only in the infinite spin-chain limit $L \gg 1$ due to usual violation of boundary conditions by the Yangian. It is natural to expect that this Yangian structure contains the Yangian of $\mathfrak{h}$ arising from the $\mathcal{S}$-matrix as a subalgebra, at least when acting on physical asymptotic states.

At the AdS end of the spectrum, which we shall focus on, the first signs of integrability in classical superstring theory on $AdS_5 \times S^5$ came with the discovery of the Bena-Polchinski-Roiban (BPR) Lax connection \cite{String Integrability}. The classical Yangian generated by the corresponding first non-local charges was identified in \cite{String Yangian}. Later on, the Poisson bracket of this Lax connection could be brought into a form exhibiting integrability \cite{Maillet}, although only in various bosonic subsectors admitting a principal chiral model representation \cite{Bosonic PCM}. Earlier attempts at bringing the Poisson structure of the full superstring on $AdS_5 \times S^5$ into the desired form had been either inconclusive \cite{attempts} or only partially satisfactory \cite{SNM}. It was only much later, after adding to the BPR Lax connection extra terms proportional to various Hamiltonian constraints (see \cite{Ham} for a justification of this extension), that this program was finally brought to completion \cite{Magro}.

The resulting classical integrable structure, valid at finite-volume, is characterised by a pair of $r/s$-matrices \cite{Maillet} which generalise the standard notion of classical $r$-matrix for so called `ultralocal' systems to the `non-ultralocal' case at hand. A remarkable, and yet quite surprising, fact is that one obtains the same kind of $r/s$-algebra with the exact same $r$- and $s$-matrices whether one works within the Green-Schwarz or the pure spinor formalism of the superstring on $AdS_5 \times S^5$ \cite{Ham, Magro}. This suggests a unified approach for quantising the superstring starting from this $r/s$-algebra.

The purpose of this paper is to reformulate the classical integrable structure of the superstring $\sigma$-model within the $R$-matrix approach \cite{R approach}, in which the pair of $r/s$-matrices admit a very natural and simple algebraic interpretation. In fact, it is by now well understood that the special property of superstrings on $AdS_5 \times S^5$ guaranteeing its classical integrability is the $\mathbb{Z}_4$-grading of its supercoset target space \cite{Berkovits:Z4}. We shall therefore work within the more general setting of $\mathbb{Z}_4$-graded supercoset $\sigma$-models and exhibit their classical integrable structure and corresponding $R$-matrix.

For conciseness we summarise the main result here. Let $\g$ be the Grassmann envelope of a $\mathbb{Z}_4$-graded Lie superalgebra $\gcirc$, with $\mathbb{Z}_4$-automorphism $\Omega$, and consider its twisted loop algebra $\Loop \gO$ in the loop variable $z$. This admits a natural vector space decomposition $\Loop \gO = \Loop \gO_+ \dotplus \Loop \gO_-$ into positive and negative powers of $z$. Introduce the standard classical $R$-matrix as the difference of projections onto these respective subspaces,
\begin{equation} \label{R intro}
R = \pi_+ - \pi_- \in \text{End } \Loop \gO. \tag{$i$}
\end{equation}
The $r$- and $s$-matrices are then given by the skew-symmetric and symmetric parts of $R$ with respect to the standard inner product on $\Loop \gO$ `twisted' by the Zhukovsky map $z \mapsto u$, namely
\begin{equation} \label{ip intro}
(X, Y)_{\phi} = \res_z \langle X(z), Y(z) \rangle du, \qquad \forall X, Y \in \Loop \gO. \tag{$ii$}
\end{equation}
Besides reproducing the correct $r/s$-matrices, the need for twisting by the Zhukovsky map can also be seen from the very special form of the Lax matrix of the supercoset $\sigma$-model,
\begin{equation} \label{Lax intro}
\L = 4 \, \phi(z)^{-1} \sum_{k=1}^{\infty} z^k \left( k A_1^{(k)} + 2 (\nabla_1 \Pi_1)^{(k)} \right), \tag{$iii$}
\end{equation}
where $\{ A_1^{(i)}, \nabla_1 \Pi_1^{(i)} \}_{i = 0}^3$ are the fields of the $\sigma$-model and $\phi(z) = z \partial_z u$. Indeed, after twisting the standard inner product by $\phi(z)$, which results in \eqref{ip intro}, the Lax matrix \eqref{Lax intro} admits a very natural description, namely it can be shown to take values in the smooth dual $(\Loop \gO_-)^{\ast}$.

As we shall show in the core of the paper, all the necessary information about the classical integrable structure of the supercoset $\sigma$-model is contained in the $R$-matrix \eqref{R intro}, the inner product \eqref{ip intro} and the Lax matrix \eqref{Lax intro}.
Indeed, the $R$-matrix \eqref{R intro} on $\Loop \gO$ lifts to the central extension $\hG$ of the current algebra $C^{\infty}(S^1, \Loop \gO)$ and the associated Poisson structure on its smooth dual $\hGd$ then corresponds exactly to the $r/s$-algebra of \cite{Magro}. Moreover, the Lax matrix \eqref{Lax intro} belongs to a certain coadjoint orbit in $\hGd$ parameterised by finitely many functions on the circle $S^1$.

\paragraph{Remarks.}
\begin{itemize}
  \item We consider the Grassmann envelope $\g$ of the Lie superalgebra $\gcirc$ for two reasons: firstly it is a Lie algebra and so the $R$-matrix approach directly applies to it, and secondly it naturally exponentiates to the Lie group with Grassmann structure $G \coloneqq \exp \g$ which is the starting point for the construction of the supercoset $\sigma$-model action.
  \item The $R$-matrix \eqref{R intro} satisfies the so-called modified classical Yang-Baxter equation, as a result of which $[X,Y]_R = \mbox{\small $\frac{1}{2}$} ([RX, Y] + [X, RY]), \forall X, Y \in \Loop \gO$ defines a second Lie bracket on $\Loop \gO$. We say that $\Loop \gO$ is a Lie \textit{di}algebra, to be distinguished from a Lie \textit{bi}algebra which admits a second Lie bracket on its \textit{dual}. When $R^{\ast} = -R$ the two structures can be identified, however due to the twist in the inner product \eqref{ip intro} this is not possible here.
  \item The $r/s$-matrices of Sch\"afer-Nameki and Mikhailov \cite{SNM} also admit an interpretation in the $R$-matrix approach, albeit a more complicated one. Specifically it should be given by a sum of $R$ in \eqref{R intro} and  a matrix $R_0$ obtained from its projection onto the grade zero part of $\g$, much like the $R$-matrix for a bosonic $\sigma$-model on a symmetric space found in \cite{Sevostyanov}.
  \item Let us also point out a possible connection with the classical $r$-matrix of Beisert-Spill \cite{Beisert-Spill}. It too presents a form of twisiting similar to \eqref{ip intro}, but is based on the symmetry algebra $\mathfrak{gl}(2|2)$ and generates a Lie bialgebra structure. Although the underlying algebras are different, it ought to be possible to relate the $R$-matrix \eqref{R intro} for $\gcirc = \mathfrak{psu}(2,2|4)$ in the large volume limit $L \gg 1$ to this classical skew-symmetric $r$-matrix which describes asymptotic scattering.
  \item The construction presented here is entirely standard \cite{R approach} apart from the twist in the inner product \eqref{ip intro}, which however was also used in \cite{Sevostyanov}. The difference here is that the twist will appear naturally as a factor in the Lax matrix \eqref{Lax intro}, expressing the need for a corresponding twist in the inner product \eqref{ip intro}. A similar factor in the Lax matrix can also be found in \cite{Maillet2}, where it was explicitly applied to the case of the principal chiral model.
  \item Since the inner product enters in the definition of the Poisson structure, the latter therefore has a natural dependence on the Zhukovsky variable $u$. This is in agreement with the earlier observation that the symplectic structure induced on the space of finite-gap solutions is canonical when expressed in the Zhukovsky variable \cite{Paper2}.
  \item The integrable structure based on \eqref{R intro}, \eqref{ip intro} and \eqref{Lax intro} does not rely on any specific properties of the Lie superalgebra $\g$ other than that it should admit a $\mathbb{Z}_4$-automorphism. The construction should therefore equally apply to all other AdS/CFT-type dualities based on $\mathbb{Z}_4$-graded Lie superalgebras \cite{otherAdSCFT}, as recently classified in \cite{Zarembo}.
  \item There is nothing special about $\mathbb{Z}_4$-gradings since it is known that Lie (super)algebras with $\mathbb{Z}_m$-gradings also give rise to actions admitting a Lax connection \cite{Young}. It would be interesting to generalise the construction of the Hamiltonian Lax connection to this setting, compute the corresponding $r/s$-matrices and interpret them in the $R$-matrix approach.
\end{itemize}

The paper is organised as follows: In section \ref{sect: 1} we introduce supercoset $\sigma$-models for a generic $\mathbb{Z}_4$-graded Lie superalgebra $\gcirc$ and review the derivation of the Lax connection in the Hamiltonian setting. In section \ref{sect: 2} we define the twisted loop algebra $\Loop \gO$ with its twisted inner product \eqref{ip intro}. In section \ref{sect: 3} we equip $\Loop \gO$ with a Lie dialgebra structure by means of the standard $R$-matrix \eqref{R intro}. We construct the current algebra $\G$ of $\Loop \gO$ in section \ref{sect: 4}, and show that its central extension $\hG$ gives rise to zero curvature type equations. In section \ref{sect: 5} we show that $\hG$ inherits a Lie dialgebra structure from $\Loop \gO$ and that the corresponding Kostant-Kirillov Poisson structure on the dual $\hGd$ is of the desired $r/s$-type with the $r/s$-matrices of \cite{Magro}. Section \ref{sect: 6} serves to reinterpret the results of \cite{Ham} and \cite{Magro} in the present formalism. We occasionally refer to appendix \ref{app: Notations} for notations.

\section{$\mathbb{Z}_4$-graded supercoset $\sigma$-models} \label{sect: 1}

\paragraph{Lie superalgebra.} Let $\gcirc$ be a Lie superalgebra admitting a $\mathbb{Z}_4$-automorphism $\Omega : \gcirc \rightarrow \gcirc$, namely such that $\Omega^4 = 1$. As a vector space, $\gcirc$ then decomposes into a direct sum $\oplus_{n=0}^3 \gcirc_n$ of eigenspaces of $\Omega$ defined by $\Omega(\gcirc_n) = i^n \gcirc_n$. We shall also assume that $\gcirc$ admits a non-degenerate, invariant, graded-symmetric, bilinear form $\langle \cdot, \cdot \rangle : \gcirc \times \gcirc \rightarrow \mathbb{C}$ which moreover respects the grading in the sense that whenever $x \in \gcirc_n$, $y \in \gcirc_m$ we have $\langle x, y \rangle = 0$ unless $(n+m) = 0$, where $(n) \coloneqq n (\text{mod } 4)$.

The Grassmann envelope $\g \coloneqq (\Gamma \otimes \gcirc)_{\bar{0}}$ of $\gcirc$ is defined as the grade zero part of its tensor product with a Grassmann algebra $\Gamma$. All the properties of $\gcirc$ extend naturally to $\g$, with the only difference being that $\langle \cdot, \cdot \rangle : \g \times \g \rightarrow \Gamma_0$ is now symmetric. Since $\g$ is a Lie algebra, one can define its corresponding Lie group (with Grassmann structure) via the exponential map $G \coloneqq \exp \g$.

\paragraph{Supercoset $\sigma$-model.} Let $H \coloneqq \exp \g_0$ denote the Lie subgroup of $G$ corresponding to the Lie subalgebra $\g_0$ of $\g$. We wish to construct a $\sigma$-model action for maps from a cylindrical worldsheet $\Sigma = S^1 \times \mathbb{R}$, parametrised by $(\sigma,\tau)$, to the supercoset $G/H$. A natural way to do this is to consider instead maps $g : \Sigma \rightarrow G$ and write down an action invariant under $g \mapsto gh$ for any $h : \Sigma \rightarrow H$ \cite{Eichenherr:1979ci}. It is also natural to require invariance under a global left $G$-action, namely $g \mapsto U g$ for $U \in G$. The 1-form $A \coloneqq - g^{-1} dg = A^{(0)} + A^{(1)} + A^{(2)} + A^{(3)} \in \Omega^1(\Sigma, \g)$ is invariant under $g \mapsto U g$ and transforms as $A \mapsto h^{-1} A h - h^{-1} dh$ under $g \mapsto g h$. Since $h^{-1} dh$ takes values in $\g_0$, it follows that the gradings $A^{(1,2,3)}$ transform homogeneously, $A^{(1,2,3)} \mapsto h^{-1} A^{(1,2,3)} h$. A possible Lagrangian is
\begin{equation} \label{Lagrangian GS}
\mathcal{L}_{GS} = - \mbox{\small $\frac{1}{2}$} \langle A^{(2)} \wedge \ast A^{(2)} \rangle - \mbox{\small $\frac{1}{2}$} \langle A^{(1)} \wedge A^{(3)} \rangle + \langle \Lambda, dA - A^2 \rangle.
\end{equation}
where for any two $\g$-valued 1-forms $\phi = \phi_a T^a, \psi = \psi_b T^b \in \Omega^1(\Sigma, \g)$, with $\{ T^a \in \g \}$ a basis of $\g$, we introduced the notation $\langle \phi \wedge \psi \rangle \coloneqq \phi_a \wedge \psi_b \langle T^a, T^b \rangle$. Besides the kinetic term $\langle A^{(2)} \wedge \ast A^{(2)} \rangle$, \eqref{Lagrangian GS} includes the usual (exact) WZ term $\langle A^{(1)} \wedge A^{(3)} \rangle$ whose coefficient is fixed to a specific value ensuring the existence of a flat connection \cite{Berkovits:Z4, Young}. Furthermore, in order to consider \eqref{Lagrangian GS} as depending on $A$ rather than $g$, the Maurer-Cartan equations $dA = A^2$ needs to be imposed using a Lagrange multiplier $\Lambda \in \g$. The resulting Lagrangian corresponds to the Green-Schwarz (GS) formulation of the superstring, or more specifically to the Metsaev-Tseytlin superstring \cite{Metsaev}.

Alternatively we could choose the natural kinetic term $\langle (A - A^{(0)}) \wedge \ast (A - A^{(0)}) \rangle$, which includes fermions as well as bosons. In this case, the total ``hybrid''-Lagrangian reads \cite{Berkovits:Z4}
\begin{equation} \label{Lagrangian PS}
\mathcal{L}_{PS} = - \mbox{\small $\frac{1}{2}$} \langle A^{(2)} \wedge \ast A^{(2)} \rangle - \langle A^{(1)} \wedge \ast A^{(3)} \rangle + \mbox{\small $\frac{1}{2}$} \langle A^{(1)} \wedge A^{(3)} \rangle + \langle \Lambda, dA - A^2 \rangle.
\end{equation}
This constitutes the matter part of the full Lagrangian in the pure-spinor (PS) formulation of the superstring. It should be supplemented with kinetic terms for the pure-spinor ghosts as well as their coupling to the gauge field $A^{(0)}$. However, due to their `wrong' statistics, ghosts do not take value in the Grassmann envelope $\g = (\Gamma \otimes \gcirc)_{\bar{0}}$ but rather in the grade one part $(\Gamma \otimes \gcirc)_{\bar{1}}$.

Although we shall not be concerned with the issue of ghosts in this paper, it is worth mentioning that the algebraic structure obtained in the absence of ghosts persists when ghosts are included in the Lagrangian \eqref{Lagrangian PS}. More precisely, the $r/s$-matrix algebra \eqref{r-s algebra} discussed below continues to hold with the exact same $r/s$-matrices \cite{Magro}.

\paragraph{Hamiltonian analysis of GS.} Going over to the Hamiltonian formalism we apply Dirac's consistency algorithm to determine all the Hamiltonian constraints. This was done for \eqref{Lagrangian GS} in \cite{Ham} with the Lie superalgebra $\gcirc = \mathfrak{psu}(2,2|4)$ in mind but the construction presented there is entirely generic. Doing so leads to a set of primary and secondary constraints, but no tertiary constraints. Subsequently, all the primary constraints as well as some secondary constraints can be done away with by imposing partial gauge fixing conditions. The resulting phase-space $\P$ is parametrised by the pair of conjugate fields $A_1, \Pi_1\in C^{\infty}(S^1, \g)$ with the canonical Poisson (Dirac) bracket
\begin{equation} \label{can Poisson bracket}
\{ A_{1\1}(\sigma), \Pi_{1\2}(\sigma') \} = C_{\1\2} \delta_{\sigma \sigma'}.
\end{equation}
This phase-space is subject to some bosonic constraints $\mathcal{T}_{\pm} \approx \C^{(0)} \approx 0$, all of which are first class, and some fermionic constraints $\C^{(1,3)} \approx 0$, which are partly second class since they fail to weakly commute among themselves $\{ \C^{(1,3)}, \C^{(1,3)} \} \not \approx 0$. The quantities $\C^{(0,1,3)} \in C^{\infty}(\P, \g)$, $\mathcal{T}_{\pm} \in C^{\infty}(\P, \mathbb{R})$ are explicitly given by
\begin{subequations} \label{constraints GS}
\begin{gather}
\label{C constraints} \C^{(0)} \coloneqq (\nabla_1 \Pi_1)^{(0)}, \qquad \C^{(1)} \coloneqq (\nabla_1 \Pi_1)^{(1)} + \mbox{\small $\frac{1}{2}$} A_1^{(1)}, \qquad \C^{(3)} \coloneqq (\nabla_1 \Pi_1)^{(3)} - \mbox{\small $\frac{1}{2}$} A_1^{(3)},\\
\label{T constraints} \mathcal{T}_+ \coloneqq \langle \A_+^{(2)}, \A_+^{(2)} \rangle - \langle A_1^{(1)}, \C^{(3)} \rangle, \qquad \mathcal{T}_- \coloneqq \langle \A_-^{(2)}, \A_-^{(2)} \rangle + \langle A_1^{(3)}, \C^{(1)} \rangle,
\end{gather}
\end{subequations}
where $\nabla_1 = \partial_{\sigma} - [A_1, \cdot]$ and $\A_{\pm}^{(2)} \coloneqq \mbox{\small $\frac{1}{2}$} ((\nabla_1 \Pi_1)^{(2)} \mp A_1^{(2)})$. Noting that the field $\Pi_1$ only ever appears in the combination $\nabla_1 \Pi_1$, we can equally parametrise the phase-space $\P$ by the pair of fields $A_1, \nabla_1 \Pi_1\in C^{\infty}(S^1, \g)$ whose Poisson brackets can be computed from \eqref{can Poisson bracket}.

The extended Hamiltonian corresponding to the Lagrangian \eqref{Lagrangian GS} is a linear combination of constraints, namely
\begin{equation} \label{ext Ham GS}
\mathcal{H}_{GS} = \rho_+ \mathcal{T}_+ + \rho_- \mathcal{T}_- - \langle \mu^{(3)}, \C^{(1)} \rangle - \langle \mu^{(1)}, \C^{(3)} \rangle - \langle \mu^{(0)}, \C^{(0)} \rangle,
\end{equation}
where the Lagrange multipliers $\rho_{\pm}$, $\mu^{(0)}$ are free. However, the requirement that $\mathcal{H}_{GS}$ be first class requires that $\mu^{(1)} \in C^{\infty}(S^1, \g_1)$ and $\mu^{(3)} \in C^{\infty}(S^1, \g_3)$ satisfy the following constraints
\begin{equation} \label{mu}
[ \A_+^{(2)}, \mu^{(1)} ] \approx [ \A_-^{(2)}, \mu^{(3)} ] \approx 0.
\end{equation}
Thus non-zero solutions of \eqref{mu} correspond to first class parts of the fermionic constraints $\C^{(1,3)}$, which in turn generate $\kappa$-symmetry. In fact, the total number of $\kappa$-symmetries reads
\begin{equation*}
\dim \ker \text{ad } \A^{(2)}_+|_{\g_1} + \dim \ker \text{ad } \A^{(2)}_-|_{\g_3}.
\end{equation*}

\paragraph{Hamiltonian analysis of PS.} Applying Dirac's procedure to the Lagrangian \eqref{Lagrangian PS} is somewhat simpler than for \eqref{Lagrangian GS}. Most of the secondary constraints are second class with primary constraints. Imposing gauge fixing conditions for the remaining primary constraints and passing to the Dirac bracket results in the same phase-space $\P$ as in \eqref{can Poisson bracket}. There are fewer constraints, $\hat{\mathcal{T}}_{\pm} \approx \C^{(0)} \approx 0$, all of which are first class and bosonic. Here $\C^{(0)}$ is as in \eqref{constraints GS} and $\hat{\mathcal{T}}_{\pm}$ are defined by
\begin{equation} \label{constraints PS}
\hat{\mathcal{T}}_{\pm} \coloneqq \mathcal{T}_{\pm} + \mbox{\small $\frac{1}{2}$} \langle \C^{(1)}, \C^{(3)} \rangle.
\end{equation}
The extended Hamiltonian corresponding to \eqref{Lagrangian PS} is a general linear combination of the first class constraints, namely
\begin{equation} \label{ext Ham PS}
\mathcal{H}_{PS} = \hat{\rho}_+ \hat{\mathcal{T}}_+ + \hat{\rho}_- \hat{\mathcal{T}}_- - \langle \hat{\mu}^{(0)}, \C^{(0)} \rangle,
\end{equation}
where the Lagrange multipliers $\hat{\rho}_{\pm} \in C^{\infty}(S^1, \mathbb{R})$ and $\hat{\mu}^{(0)} \in C^{\infty}(S^1, \g_0)$ are unconstrained.

\paragraph{Lax connection.} In any 2-d field theory, a Lax connection enables one, by virtue of its flatness, to generate integrals of motion, which in turn should generate symmetries via the Poisson bracket. If the system happens to be constrained, then the integrals all ought to be first class in order to preserve the constraint surface. This requirement alone turns out to be almost enough to construct the (spatial component of a) Lax connection \cite{Ham}. Indeed, considering a general linear combination of the phase-space variables $\L = A_1^{(0)} + \rho \, \C^{(0)} + \sum_{i=1}^3 (a_i A_1^{(i)} + b_i (\nabla_1 \Pi_1)^{(i)})$, then in both the GS and PS formalisms this fixes all coefficients except $\rho$ in terms of a single parameter $z \in \mathbb{C}$. Note that since $\C^{(0)} \approx 0$, this means $\L$ is already entirely determined on the constraint surface.

To fix $\rho$ we demand that $\L$ also be the spatial component of a flat connection \emph{off} the constraint surface. This requires defining the generators of $\sigma$- and $\tau$-translations on the whole of $\P$, which in both the GS and PS formalisms we take to be
\begin{equation}
\P_1 \coloneqq \hat{\mathcal{T}}_+ - \hat{\mathcal{T}}_- - \langle A_1^{(0)}, \C^{(0)} \rangle, \qquad \P_0 \coloneqq \hat{\mathcal{T}}_+ + \hat{\mathcal{T}}_- - \langle A_1^{(0)}, \C^{(0)} \rangle.
\end{equation}
Notice that the issue of $\tau$-translations is more delicate since $\tau$ is not intrinsically defined as opposed to $\sigma$. One can, for instance, add a term like $\langle \C^{(1)}, \C^{(3)} \rangle$ in \eqref{constraints PS} which is quadratic in the constraints since this does not affect the equations of motion on the constraint surface.

It is then a matter of determining $\rho$, as a function of $z$, such that there exists an $\M$ with
\begin{equation*}
\{ \L, P_0 \} = \partial_{\sigma} \M + [\M, \L],
\end{equation*}
where $P_0 \coloneqq \int d\sigma \, \P_0(\sigma)$. Note that this is a strong equation, valid in the whole phase-space $\P$ and not just on the constraint surface. It is convenient to write the Lax connection $(\L, \M)$ in light-cone components $(\L_+, \L_-)$ and in terms of which $\L = \L_+ - \L_-$. Explicitly it reads \cite{Ham}
\begin{subequations} \label{light-cone L}
\begin{align}
\L_+ &\coloneqq A_1^{(0)} + \frac{1}{2} (\nabla_1 \Pi_1)^{(0)} + \sum_{j=1}^4 z^j A_1^{(j)} - \frac{1}{4} \left( \sum_{j=1}^4 j z^j A_1^{(j)} + 2 \sum_{j=1}^4 z^j (\nabla_1 \Pi_1)^{(j)} \right),\\
\L_- &\coloneqq A_1^{(0)} + \frac{1}{2} (\nabla_1 \Pi_1)^{(0)} - \frac{1}{4 z^4} \left( \sum_{j=1}^4 j z^j A_1^{(j)} + 2 \sum_{j=1}^4 z^j (\nabla_1 \Pi_1)^{(j)} \right),
\end{align}
\end{subequations}
where superscripts in brackets are always taken $\text{mod } 4$, \text{e.g.} $A_1^{(4)} \equiv A_1^{(0)}$ and $(\nabla_1 \Pi_1)^{(4)} \equiv (\nabla_1 \Pi_1)^{(0)}$. Note that $z \mapsto z^{-1}$ leads to an alternative parametrisation of this Lax connection.

\paragraph{$r/s$-matrix algebra.} To further guarantee the involution of the integrals of motion generated by a Lax connection, the bracket of its spatial component $\L$ should take the generic form \cite{Maillet}
\begin{equation} \label{r-s algebra}
\{ \L_{\1} , \L_{\2} \} = [r_{\1\2} - s_{\1\2}, \L_{\1}] \delta_{\sigma_1 \sigma_2} + [r_{\1\2} + s_{\1\2}, \L_{\2} ] \delta_{\sigma_1 \sigma_2} - 2 s_{\1\2} \delta'_{\sigma_1 \sigma_2},
\end{equation}
where the tensor indices $\1$ and $\2$ respectively imply a dependence on $(\sigma, z)$ and $(\sigma', z')$. In the case of $\L = \L_+ - \L_-$ with $\L_{\pm}$ given in \eqref{light-cone L}, the $r/s$-matrices were computed in \cite{Magro}. Although this computation was performed in the context of the Lie superalgebra $\gcirc = \mathfrak{psu}(2,2|4)$, it is quite general and equally applies to the case of a generic $\mathbb{Z}_4$-graded Lie superalgebra $\gcirc$.

\section{Twisted loop algebra with twisted inner product} \label{sect: 2}

\paragraph{Loop algebra.} In view of \eqref{light-cone L}, it is natural to consider the loop algebra $\Loop \g \coloneqq \g \bb{z,z^{-1}}$ of formal Laurent series in $z$ (with a finite but arbitrary number of negative terms). Let us denote by $\Loop \g_+ \coloneqq \g \bb{z}$ the Lie subalgebra of formal Taylor series in $z$ and by $\Loop \g_- \coloneqq z^{-1} \g [z^{-1}]$ that of polynomials in $z^{-1}$ without constant term, so that $\L_{\pm} \in C^{\infty}(\P, \Loop \g_{\pm})$.
In this formalism, the transformation $z \mapsto z^{-1}$ simply corresponds to a reparametrisation of the loop algebra $\Loop \g$.

Given $X(z) \in \Loop \g$, define $\pi_- X(z) \in \Loop \g_-$ to be its pole part and $\pi_+ X(z) \in \Loop \g_+$ its regular part. It follows that $\pi_-$ and $\pi_+ = \text{id} - \pi_-$ are complementary projections and moreover $\Loop \g$ decomposes, as a vector space, into a direct sum of two Lie subalgebras
\begin{equation} \label{decomposition}
\Loop \g = \Loop \g_+ \dotplus \Loop \g_-.
\end{equation}

\paragraph{Let's do the twist.} The action of the $\mathbb{Z}_4$-automorphism $\Omega : \g \rightarrow \g$ on the Lax connection \eqref{light-cone L} is simply $\Omega(\L_{\pm}(z)) = \L_{\pm}(i z)$. This suggests extending $\Omega$ to $\Loop \g$ by setting
\begin{equation*}
\hat{\Omega} : \Loop \g \rightarrow \Loop \g, \quad \hat{\Omega}(X)(z) = \Omega(X(- i z))
\end{equation*}
so that $\hat{\Omega}(\L_{\pm}) = \L_{\pm}$. Therefore the components of the Lax connection take values in the stable subalgebra $\Loop \gO \coloneqq \{ X \in \Loop \g \,|\, \hat{\Omega}(X) = X \}$ under the action of $\hat{\Omega}$, called the twisted loop algebra. Setting $\Loop \gO_{\pm} = \Loop \gO \cap \Loop \g_{\pm}$ we have in particular $\L_{\pm} \in C^{\infty}(\P, \Loop \gO_{\pm})$ and \eqref{decomposition} implies
\begin{equation} \label{decomposition Omega}
\Loop \gO = \Loop \gO_+ \dotplus \Loop \gO_-.
\end{equation}
Concretely, $\Loop \gO = \oplus_{n=-\infty}^{\infty} \, \g_{(n)} \cdot z^n \subset \Loop \g$ and the subalgebras $\Loop \gO_{\pm}$ can be explicitly written as
\begin{equation} \label{LgO pm}
\Loop \gO_+ = \oplus_{n \geq 0} \, \g_{(n)} \cdot z^n, \qquad \Loop \gO_- = \oplus_{n < 0} \, \g_{(n)} \cdot z^n,
\end{equation}
where it is understood that elements of $\Loop \gO_-$ may contain an arbitrary but finite number of terms.

\paragraph{Come on let's twist again.} The standard inner-product on $\Loop \gO$ reads
\begin{equation} \label{standard ip}
(X, Y) \coloneqq \oint \frac{dz}{2 \pi i z} \langle X(z), Y(z) \rangle,
\end{equation}
where $\oint$ picks out the coefficient of $z^{-1}$ in the formal Laurent series. Note that since $\langle X(z), Y(z) \rangle$ is a Laurent expansion in $z^4$, the factor of $z^{-1}$ in the measure is required to pick out a non-zero coefficient, in this case its constant term. We shall, however, need to `twist' this inner-product by a given function of $z^4$, namely
\begin{equation} \label{twisted ip}
(X, Y)_{\phi} \coloneqq \oint \frac{dz}{2 \pi i z} \phi(z) \langle X(z), Y(z) \rangle, \qquad \phi(z) \coloneqq \frac{16 z^4}{(1 - z^4)^2}.
\end{equation}
The measure $\phi(z) z^{-1} dz$ is none other than $du$ where $u = 2 \frac{1+z^4}{1-z^4}$ is the Zhukovsky variable, \textit{i.e.}
\begin{equation*}
(X,Y)_{\phi} = \oint \frac{du}{2 \pi i} \langle X(z), Y(z) \rangle.
\end{equation*}
Moreover, since the integrand $\langle X(z), Y(z) \rangle$ is a Laurent expansion in $z^4$ it can also be rewritten in terms of $u$. The contour integral defining the twisted inner-product \eqref{twisted ip} therefore makes sense in the $u$-plane, even though the original twisted loop algebra $\Loop \gO$ is written in the $z$-variable.

\paragraph{Smooth dual.} The non-degenerate bilinear pairing \eqref{twisted ip} provides an embedding of $\Loop \gO$ into its algebraic dual, whose image $(\Loop \gO)^{\ast} \coloneqq (\Loop \gO, \cdot )_{\phi}$ is called the smooth dual.

It is straightforward to see that the subspaces orthogonal to \eqref{LgO pm} with respect to the standard inner product \eqref{standard ip} are, respectively
\begin{equation} \label{ortho subs}
(\Loop \gO_+)^{\perp} = \oplus_{n > 0} \, \g_{(n)} \cdot z^n, \qquad (\Loop \gO_-)^{\perp} = \oplus_{n \leq 0} \, \g_{(n)} \cdot z^n,
\end{equation}
where once again elements of $(\Loop \gO_-)^{\perp}$ can have only a finite but arbitrary number of terms.
Since \eqref{standard ip} is non-degenerate, the smooth duals of the spaces $\Loop \gO_{\pm}$ with respect to this inner-product are, respectively, $(\Loop \gO_{\mp})^{\perp}$. Then with respect to the twisted inner-product \eqref{twisted ip} we have
\begin{equation} \label{smooth dual}
(\Loop \gO_-)^{\ast} \simeq \phi^{-1} (\Loop \gO_+)^{\perp} \coloneqq \{ \phi^{-1} X \, | \, X \in (\Loop \gO_+)^{\perp} \} \subset \Loop \gO,
\end{equation}
where the last inclusion follows because $\phi(z)^{-1} = \frac{1}{16} z^4 - \frac{1}{8} + \frac{1}{16} z^{-4}$ is a Laurent polynomial in $z^4$.

\paragraph{Lax matrix.} We have seen that both components of the Lax connection \eqref{light-cone L} belong to the twisted loop algebra, namely $\L_{\pm} \in C^{\infty}(S^1, \Loop \gO)$. Of particular interest in the following will be its spatial component $\L = \L_+ - \L_-$ which using \eqref{light-cone L} explicitly reads
\begin{equation} \label{spatial L}
\L = \sum_{j=1}^4 z^j A_1^{(j)} + \frac{1 - z^4}{4 z^4} \left( \sum_{j=1}^4 j z^j A_1^{(j)} + 2 \sum_{j=1}^4 z^j (\nabla_1 \Pi_1)^{(j)} \right).
\end{equation}

\begin{lemma}
For any $b = \sum_{i = 0}^3 b^{(i)} \in \g$ we have, formally,
\begin{equation*}
\sum_{j=1}^4 z^j b^{(j)} = (1 - z^4) \sum_{k=1}^{\infty} z^k b^{(k)}, \qquad
\sum_{j=1}^4 z^j b^{(j)} + \frac{1 - z^4}{4 z^4} \sum_{j=1}^4 j z^j b^{(j)} = \frac{(1 - z^4)^2}{4 z^4} \sum_{k=1}^{\infty} k z^k b^{(k)}.
\end{equation*}
\end{lemma}

Using these relations we are able to write the Lax matrix \eqref{spatial L} in a more transparent form
\begin{equation} \label{Lax matrix}
\L = 4 \, \phi(z)^{-1} \sum_{k=1}^{\infty} z^k \left( k A_1^{(k)} + 2 (\nabla_1 \Pi_1)^{(k)} \right).
\end{equation}
Note that the factor of $\phi(z)^{-1}$ appears naturally from the Lax connection \eqref{light-cone L} itself, effectively justifying the need for the twist in the inner-product \eqref{twisted ip}. Indeed, we can immediately infer from \eqref{Lax matrix} that the Lax matrix naturally takes values in the dual space \eqref{smooth dual}, namely
\begin{equation} \label{L in dual}
\L \in C^{\infty}(\P, (\Loop \gO_-)^{\ast}).
\end{equation}
Let us close this section by noting that although the sum in \eqref{Lax matrix} is infinite, all but the first eight terms are redundant. Indeed, one can extract $A_1$ and $\nabla_1 \Pi_1$ entirely from the first eight terms.

\section{Classical $R$-matrix and mCYBE} \label{sect: 3}

In this section we endow the Lie algebra $\Loop \g$ (and hence also $\Loop \gO$) with a Lie dialgebra structure, namely a second Lie bracket. This is done by introducing an $R$-matrix $R \in \text{End } \Loop \g$ satisfying the modified Yang-Baxter equation \cite{R approach}, which allows one to define the $R$-bracket,
\begin{equation} \label{R-bracket Lg}
[X,Y]_R \coloneqq \mbox{\small $\frac{1}{2}$} \left([RX, Y] + [X, RY] \right).
\end{equation}

\paragraph{The $R$-matrix.} The most basic example of an $R$-matrix is provided by the Adler-Kostant-Symes scheme, which relies on the existence of a splitting of the underlying Lie algebra $\Loop \g$ into a (vector space) direct sum of two subalgebras, such as in \eqref{decomposition}. The corresponding standard $R$-matrix with respect to this decomposition reads
\begin{equation} \label{R standard}
R \coloneqq \pi_+ - \pi_-.
\end{equation}
Note that this is skew-symmetric with respect to the untwisted inner product on $\Loop \g$,
\begin{equation} \label{ip}
(X,Y)' \coloneqq \oint \frac{dz}{2 \pi i} \langle X(z), Y(z) \rangle.
\end{equation}
Indeed, $(RX, Y)' = -  (X, RY)'$ follows immediately from observing that $(\pi_+ X, Y)' = (X, \pi_- Y)'$. Let $R^{\ast}$ be the adjoint with respect to the twisted inner product \eqref{twisted ip}, \textit{i.e.} $(RX, Y)_{\phi} = (X, R^{\ast} Y)_{\phi}$ or equivalently $(RX, \varphi Y)' = (X, \varphi R^{\ast} Y)'$ where $\varphi(z) \coloneqq \frac{du}{dz} = \phi(z) z^{-1}$. It then follows that
\begin{equation} \label{adj R}
R^{\ast} = - \varphi^{-1} R \varphi.
\end{equation}
Before proceeding, let us first determine the tensor kernel (see appendix \ref{app: Notations}) of the $R$-matrix \eqref{R standard} with respect to the twisted inner product \eqref{twisted ip}.

\paragraph{Projection kernels.} It is straightforward to check that with respect to the inner product \eqref{twisted ip}, the kernels of the pair of projections $\pi_{\pm} : \Loop \g \rightarrow \Loop \g_{\pm}$ explicitly read
\begin{equation*}
\pi_{- \1\2}(z_1, z_2) = \sum_{m=1}^{\infty} \left(\! \frac{z_2}{z_1} \!\right)^{\!m} \!C_{\1\2}^{(-m \; m)} \phi(z_2)^{-1}, \qquad \pi_{+ \1\2}(z_1, z_2) = \sum_{m=0}^{\infty} \left(\! \frac{z_1}{z_2} \!\right)^{\!m} \!C_{\1\2}^{(m \; -m)} \phi(z_2)^{-1},
\end{equation*}
in the sense that $(\pi_{\pm \1\2}(z_1, z_2), X_{\2}(z_2))_{\phi \2} = (\pi_{\pm} X)_{\1}(z_1)$ and where $C^{(i \, j)}_{\1\2}$ is the component of the tensor Casimir along $\gcirc_{(i)} \otimes \gcirc_{(j)}$, see appendix \ref{app: Notations}. Both geometric progressions can be summed up provided $|z_2| < |z_1|$ in the case of $\pi_-$ and $|z_1| < |z_2|$ for $\pi_+$. We can then write
\begin{subequations} \label{proj kern resum twist}
\begin{alignat}{2}
\label{proj kern resum twist a} \pi_{- \1\2}(z_1, z_2) &= \frac{\sum_{j=0}^3 z_1^j z_2^{4-j} C^{(j\, 4-j)}_{\1\2}}{z_1^4 - z_2^4} \phi(z_2)^{-1}, \qquad && |z_2| < |z_1|,\\
\label{proj kern resum twist b} \pi_{+ \1\2}(z_1, z_2) &= \frac{\sum_{j=0}^3 z_1^j z_2^{4-j} C^{(j\, 4-j)}_{\1\2}}{z_2^4 - z_1^4} \phi(z_2)^{-1}, \qquad && |z_2| > |z_1|.
\end{alignat}
\end{subequations}

\begin{figure}
\centering
\psfrag{z1}{$z_1$} \psfrag{z2}{$- i z_1$} \psfrag{z3}{$- z_1$} \psfrag{z4}{$i z_1$}
\psfrag{Gp}{$\Gamma_+$} \psfrag{Gm}{$\Gamma_-$} \psfrag{G}{$\Gamma$}
\begin{tabular}{ccc}
\includegraphics[height=40mm]{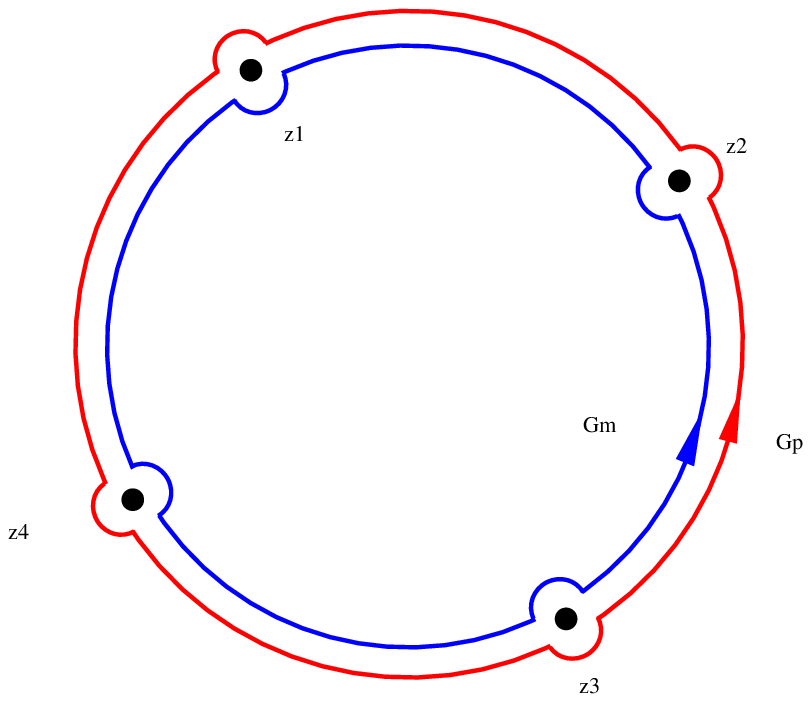} & $\qquad \qquad$ & \includegraphics[height=40mm]{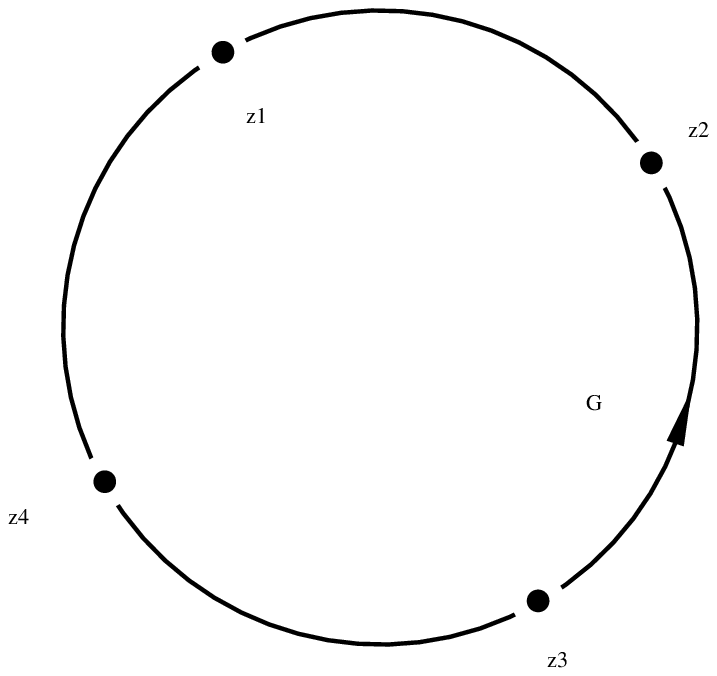}\\
\\
$(a)$ & & $(b)$
\end{tabular}
\caption{The black dots denote the poles of the projection kernels \eqref{proj kern resum twist} at $i^n z_1$ in the $z_2$-plane. Figure $(a)$ shows the (red) contour $\Gamma_+$ for $\pi_+$ which contains all four singularities and the (blue) contour $\Gamma_-$ for $\pi_-$ which goes around the origin but  contains none of the singularities. Figure $(b)$ shows the contour for the $R$-matrix, namely $\Gamma = \{ |z_2| = |z_1| \}$ with all four singularities cut out.}
\label{fig: R-matrix}
\end{figure}

\paragraph{$R$-matrix kernel.} We would like to take the difference of the expressions in \eqref{proj kern resum twist} to obtain a simple expression for the kernel of $R$, but unfortunately their domains of validity do not overlap. To overcome this difficulty we note that the projections $\pi_{\pm}$ can be written as contour integrals of the common kernel
\begin{equation*}
\kappa_{\1\2}(z_1,z_2) \coloneqq \frac{\sum_{j=0}^3 z_1^j z_2^{4-j} C^{(j\, 4-j)}_{\1\2}}{z_2^4 - z_1^4} \phi(z_2)^{-1}
\end{equation*}
around contours $\Gamma_{\pm}$ which are entirely contained in the regions $|z_2| > |z_1|$ and $|z_2| < |z_1|$ respectively, as depicted in Figure \ref{fig: R-matrix}$(a)$,
namely
\begin{equation} \label{proj kern int}
(\pi_{\pm} X)_{\1}(z_1) = \pm \oint_{\Gamma_{\pm}} \frac{dz_2}{2 \pi i z_2} \phi(z_2) \langle \kappa_{\1\2}(z_1, z_2), X_{\2}(z_2) \rangle_{\2}.
\end{equation}
We can thus also write the action of \eqref{R standard} as a contour integral
\begin{equation*}
(R X)_{\1}(z_1) = \left( \oint_{\Gamma_+} + \oint_{\Gamma_-} \right) \frac{dz_2}{2 \pi i z_2} \phi(z_2) \langle \kappa_{\1\2}(z_1, z_2), X_{\2}(z_2) \rangle_{\2}.
\end{equation*}
In the limit where the contours $\Gamma_{\pm}$ converge to $\Gamma \coloneqq \{ |z_2| = |z_1| \}$, the contributions from the small semi-circles around the singularities at $i^n z_1$ cancel (since each contribute opposite half residues) and we are left with twice the principal value of the integral around $\Gamma$, namely
\begin{equation} \label{R kernel def}
(R X)_{\1}(z_1) = 2\, \text{v.p.} \oint_{\Gamma} \frac{dz_2}{2 \pi i z_2} \phi(z_2) \langle \kappa_{\1\2}(z_1, z_2), X_{\2}(z_2) \rangle_{\2},
\end{equation}
In conclusion the kernel of the $R$-matrix admits the following expression,
\begin{subequations} \label{proj kern resum twist 2}
\begin{equation} \label{proj kern resum twist 2a}
R_{\1\2}(z_1, z_2) = 2\, \text{v.p.} \frac{\sum_{j=0}^3 z_1^j z_2^{4-j} C^{(j\, 4-j)}_{\1\2}}{z_2^4 - z_1^4} \phi(z_2)^{-1},
\end{equation}
where the principal value symbol is to be understood in the sense of the preceding integral \eqref{R kernel def}, along the contour shown in Figure \ref{fig: R-matrix}$(b)$. Likewise we obtain from \eqref{adj R}, or more directly using $R^{\ast}_{\1\2} = P(R_{\1\2})$ (see appendix \ref{app: Notations}), an expression for the kernel of the adjoint $R^{\ast}$, namely
\begin{equation} \label{proj kern resum twist 2b}
R^{\ast}_{\1\2}(z_1, z_2) = 2\, \text{v.p.} \frac{\sum_{j=0}^3 z_1^{4-j} z_2^j C^{(4-j \, j)}_{\1\2}}{z_1^4 - z_2^4} \phi(z_1)^{-1}.
\end{equation}
\end{subequations}

\paragraph{Yang-Baxter equation.} It is straightforward to check, using the fact that $\Loop \g_{\pm} \subset \Loop \g$ are Lie subalgebras, that \eqref{R standard} satisfies the modified classical Yang-Baxter equation (mCYBE),
\begin{equation} \label{mCYBE}
[RX, RY] - R([RX, Y] + [X, RY]) = - [X,Y], \qquad \forall X,Y \in \Loop \g.
\end{equation}
Moreover, this is a sufficient condition for \eqref{R-bracket Lg} to satisfy the Jacobi identity.

Taking the inner product with an arbitrary $Z \in \Loop \g$, its right hand side defines a 3-form on $\Loop \g$, namely $\hat{\omega}(X,Y,Z) \coloneqq ([X,Y],Z)_{\phi}$. Let $\hat{\omega}_{\1\2\3}$ denote its tensor kernel with respect to \eqref{twisted ip} which can be written explicitly as
\begin{equation} \label{omega}
\hat{\omega}_{\1\2\3}(z_1,z_2,z_3) \coloneqq \omega_{\1\2\3} \, \phi(z_1)^{-2} \delta(z_1 - z_2) \delta(z_2 - z_3),
\end{equation}
where $\omega_{\1\2\3}$ is the tensor of the 3-form on $\g$ given by $\omega(x,y,z) \coloneqq \langle [x,y], z \rangle$ for $x,y,z \in \g$.

Using this to express the mCYBE \eqref{mCYBE} in tensor notation we obtain
\begin{equation} \label{mCYBE tens}
[R_{\1\2}, R_{\1\3}] + [R_{\1\2}, R_{\2\3}] + [R_{\3\2}, R_{\1\3}] = - \hat{\omega}_{\1\2\3}.
\end{equation}
Despite appearances, this is not quite the same as the usual modified classical Yang-Baxter equation on the $r$-matrix in the Lie bialgebra setting, which we shall call the $r$mCYBE,
\begin{equation} \label{rmCYBE tens}
[r_{\tensor{12}}, r_{\tensor{13}}] + [r_{\tensor{12}}, r_{\tensor{23}}] + [r_{\tensor{13}}, r_{\tensor{23}}] = - \hat{\omega}_{\1\2\3}.
\end{equation}
The difference between \eqref{mCYBE tens} and \eqref{rmCYBE tens} can be understood by recalling (see appendix \ref{app: Notations}) that the kernel $R^{\ast}_{\tensor{12}}$ of $R^{\ast}$ was $P(R_{\1\2})$, in other words $R^{\ast}_{\tensor{12}} = R_{\tensor{21}}$. Therefore rewriting \eqref{mCYBE tens} as
\begin{equation} \label{mCYBE tens 2}
[R_{\tensor{12}}, R_{\tensor{13}}] + [R_{\tensor{12}}, R_{\tensor{23}}] + [R_{\tensor{13}}, (-R^{\ast})_{\tensor{23}}] = - \hat{\omega}_{\1\2\3}
\end{equation}
it becomes clear that if $R^{\ast} = - R$ then the $R$-matrix satisfies the usual $r$mCYBE. In fact one can show that when $R$ is assumed skew-symmetric, the $R$-matrix construction yields a coboundary Lie bialgebra with $R$ as $r$-matrix. Indeed, suppose $\h$ is a coboundary Lie bialgebra with 1-cocycle $\delta : \h \rightarrow \h \wedge \h$ given by $\delta(X) = \mbox{\small $\frac{1}{2}$}[1 \otimes X + X \otimes 1, r]$, $X \in \h$. Its dual Lie bracket on $\h^{\ast}$ then reads
\begin{equation*}
[\xi, \xi']_{\ast} = \mbox{\small $\frac{1}{2}$} \left( ad^{\ast} r \xi \cdot \xi' + ad^{\ast} r^{\ast} \xi' \cdot \xi \right), \qquad \forall \xi, \xi' \in \h^{\ast}.
\end{equation*}
Under the identification $\h^{\ast} \simeq \h$ and assuming $r^{\ast} = - r$, this is exactly of the form \eqref{R-bracket Lg} with $R = r$.
Yet in the case at hand the $R$-matrix is \textit{not} skew-symmetric but instead satisfies \eqref{adj R}, due to the twist in the inner product \eqref{twisted ip}.

\section{Centrally extended current algebra} \label{sect: 4}

Recall that the phase-space $\P$ is parametrised by a pair of fields in $C^{\infty}(S^1, \g)$, which is a particular realisation of the loop algebra over $\g$, the so-called current algebra.
In the previous section we had considered a different kind of realisation, namely the twisted loop algebra $\Loop \gO$, and equipped it with a Lie dialgebra structure. We can anticipate that studying the integrable structure of $\P$ will require working with a double loop algebra over $\g$, specifically the current algebra of $\Loop \gO$.

In this section we show in fact that in order to obtain zero curvature equations \---\ which constitute the main ingredient in any $2$-d integrable field theory \---\ the relevant algebra to consider is a central extension of the current algebra of $\Loop \gO$.

\paragraph{Double loop algebra.} Consider the current algebra $\G \coloneqq C^{\infty}(S^1,\Loop \gO)$ of smooth loops in $\Loop \gO$. Introducing the subalgebras $\G_{\pm} \coloneqq C^{\infty}(S^1,\Loop \gO_{\pm})$, the decomposition \eqref{decomposition} leads to
\begin{equation} \label{G decomp}
\G = \G_+ \dotplus \G_-.
\end{equation}
Let us also introduce the subalgebras $\G_{\pm}^{\perp} \coloneqq C^{\infty}(S^1, (\Loop \gO_{\pm})^{\perp})$ corresponding to \eqref{ortho subs}. We define a non-degenerate, invariant, bilinear form on $\G$, given for $\X,\Y \in \G$ by
\begin{equation} \label{ip on G}
(\!( \X, \Y )\!)_{\phi} \coloneqq \int_{S^1} d\sigma (\X(\sigma),\Y(\sigma))_{\phi}.
\end{equation}
It gives an embedding of $\G$ into its dual $\G'$ whose image $\Gd \subset \G'$ defines the smooth dual. Owing to the twist in the inner product, the smooth dual can be identified as
\begin{equation} \label{smooth dual G}
\Gd \simeq \phi^{-1} \G \coloneqq \{ \phi^{-1} \X \, | \, \X \in \G \} \subset \G.
\end{equation}

\paragraph{Central extension.} The current algebra $\G$ admits an important central extension $\hG$ defined by the 2-cocycle
\begin{equation} \label{2-cocycle}
\omega(\X,\Y) \coloneqq \int_{S^1} d\sigma (\X(\sigma),\partial_{\sigma} \Y(\sigma))_{\phi}.
\end{equation}
Specifically, $\hG$ is identified as a vector space with $\G \oplus \mathbb{C}$ and equipped with the Lie bracket
\begin{equation} \label{bracket extension}
[(\X,a), (\Y,b)] \coloneqq ([\X,\Y], \omega(\X,\Y)).
\end{equation}
Notice that this definition doesn't depend on $a$ or $b$, corresponding to the fact that the extension is indeed central. Extending the inner product \eqref{ip on G} to $\hG$ as
\begin{equation} \label{ip on hG}
(\!( (\X,a), (\Y,b) )\!)_{\phi} \coloneqq (\!( \X, \Y )\!)_{\phi} + a b,
\end{equation}
one can identify the smooth dual $\hGd$ with $\phi^{-1} \G \oplus \mathbb{C}$.

\paragraph{Coadjoint action.} The point of centrally extending $\G$ by \eqref{2-cocycle} can be understood by looking at the coadjoint action of $\hG$ on $\hGd$ defined as
\begin{equation} \label{coadj action}
(\!( ad^{\ast} (\M,c) \cdot (\X,a), (\Y, b) )\!)_{\phi} \coloneqq - (\!( (\X,a), [(\M,c),(\Y,b)] )\!)_{\phi}.
\end{equation}
Notice that the right hand side does not depend on $c \in \mathbb{C}$, using \eqref{bracket extension}. In other words, the center of $\hG$ acts trivially in the coadjoint representation and \eqref{coadj action} equally defines the coadjoint action of $\G$ on $\hGd$. It is easy to show using \eqref{2-cocycle}, \eqref{bracket extension} and \eqref{ip on hG} that the latter takes the form
\begin{equation*}
ad^{\ast} \M \cdot (\X,a) = ([\M,\X] + a\, \partial_{\sigma} \M, 0).
\end{equation*}
Observing that $a \in \mathbb{C}$ is invariant under this action we are free to set $a = 1$, which amounts to restricting attention to the subspace $\hGd_1 = \phi^{-1} \G \oplus \{1\} \subset \phi^{-1} \G \oplus \mathbb{C}$, isomorphic to $\Gd$. Given any $\X \in \G$ (respectively $\X \in \Gd$), we shall use the notation $\hat{\X} \coloneqq (\X, 1)$ for the corresponding element in $\hG_1$ (respectively $\hGd_1$). Now although we can effectively ignore the central extension since $\hGd_1$ is isomorphic to $\Gd$, the extra $\sigma$-derivative term in the coadjoint action of $\G$ on $\hGd_1$, namely
\begin{equation} \label{coadjoint central}
ad^{\ast} \M \cdot \hat{\X} = ([\M,\X] + \partial_{\sigma} \M, 0)
\end{equation}
and which is absent in the usual coadjoint action of $\G$ on $\Gd$, is a remnant of this central extension.

\paragraph{The Kostant-Kirillov bracket.} The Lie algebra structure \eqref{bracket extension} on $\hG$ induces a natural Poisson structure on $\hGd_1$. Given $f \in C^{\infty}(\hGd)$ we define its derivative $\hat{d}f = (df, c_f) \in \hG$ at $(\L,a) \in \hGd$ as
\begin{equation} \label{Frechet der}
(\!( (\X,b), \hat{d}f )\!)_{\phi} \coloneqq \mbox{\small $\left. \frac{d}{dt} \right|_{t=0}$} f( (\L,a) + t (\X,b)), \qquad \forall (\X,b) \in \hGd.
\end{equation}
Then for any two functions $f,g \in C^{\infty}(\hGd_1)$ we set
\begin{equation} \label{Kostant-Kirillov}
\{ f, g \}(\hat{\L}) \coloneqq (\!( \hat{\L}, [ \hat{d}f, \hat{d}g ] )\!)_{\phi}.
\end{equation}
The anti-symmetry of \eqref{Kostant-Kirillov} and the Jacobi identity follow from that of the Lie bracket on $\hG$. This defines the Kostant-Kirillov bracket, making $\hGd_1$ into a Poisson manifold. Although the Poisson bracket \eqref{Kostant-Kirillov} is very natural indeed, it is important to note that it is degenerate. Its kernel is easily shown to coincide with the set $\mathcal{I}(\hGd_1) \subset C^{\infty}(\hGd_1)$ of coadjoint invariant functions on $\hGd_1$, which itself is characterised by
\begin{equation} \label{coadj inv property}
f \in \mathcal{I}(\hGd_1) \quad \Longleftrightarrow \quad ad^{\ast} df \cdot \hat{\L} = 0, \; \forall \hat{\L} \in \hGd_1.
\end{equation}

Now consider the simplest type of function on $\hGd_1$, namely a linear one. It can be specified by an element $\hat{\X} \in (\hGd_1)^{\ast} \simeq \hG_1$ and is given explicitly as $\hat{\X} : \hat{\L} \mapsto (\!( \hat{\L}, \hat{\X} )\!)_{\phi}$. From the definition \eqref{Frechet der} we find $\hat{d} \hat{\X} = \hat{\X}$. Its evolution under the flow $\partial_t \coloneqq \{ \cdot, h \}$ corresponding to some $h \in C^{\infty}(\hGd_1)$ reads
\begin{equation*}
\partial_t (\!( \hat{\L}, \hat{\X} )\!)_{\phi} = \{ \hat{\X}, h \}(\hat{\L}) = (\!( ad^{\ast} dh \cdot \hat{\L}, \hat{\X} )\!)_{\phi}.
\end{equation*}
In other words, the flow of any $h \in C^{\infty}(\hGd_1)$ under the Kostant-Kirillov bracket \eqref{Kostant-Kirillov} is tangent to the coadjoint orbit of $\G$ in $\hGd_1$ since
\begin{equation} \label{coadj orbit}
\partial_t \hat{\L} = ad^{\ast} dh \cdot \hat{\L}.
\end{equation}
Making use of \eqref{coadjoint central} this takes the form of a zero curvature equation,
\begin{equation} \label{pre Lax}
\partial_t \L - \partial_{\sigma} \N = [\N, \L], \qquad \text{where} \quad \N = dh.
\end{equation}
One might hastily conclude that the space $\hGd_1$ provides a natural setting for describing integrable $2$-d field theories. The problem is that \emph{any} function $h \in C^{\infty}(\hGd_1)$ generates a zero curvature flow, which is unlikely to yield a valid setting for non-trivial examples of integrable $2$-d field theories. Indeed, suppose that $f \in C^{\infty}(\hGd_1)$ is invariant under all these flows, that is $0 = \partial_t f = \{ f, h \}$, then necessarily it must lie in the kernel of the Kostant-Kirillov bracket \eqref{Kostant-Kirillov}. In other words, such an $f$ generates a \emph{trivial} flow on $\hGd_1$.

\section{Kostant-Kirillov $R$-bracket} \label{sect: 5}

\paragraph{The $\bm{R}$-bracket.} In order to circumvent this difficulty one is led to the introduction of a second Poisson structure on $\hGd_1$, such that the Casimirs of \eqref{Kostant-Kirillov} are not in its center. One way to achieve this is to endow $\hG$ itself with a second Lie bracket distinct from the original one. In the $R$-matrix approach this second bracket is constructed from a given linear operator $R \in \text{End }\hG$. Starting from $R \in \text{End}\; \Loop \gO$ given in \eqref{R standard} we can extend it to $\text{End}\; \G$ trivially by setting $(R\X)(\sigma) = R (\X(\sigma))$, which allows us to define the $R$-bracket on $\G$ by the usual formula
\begin{equation} \label{R-bracket}
[ \X, \Y]_R \coloneqq \mbox{\small $\frac{1}{2}$} \left( [R \X, \Y] + [\X, R \Y] \right).
\end{equation}
The anti-symmetry is evident and the Jacobi identity holds by virtue of the mCYBE \eqref{mCYBE}. One sometimes denotes vector space $\G$ equipped with the $R$-bracket \eqref{R-bracket} as $\G_R$, to distinguish it from the Lie algebra $\G$ with the original Lie bracket.

We then further extend the $R$-matrix to $\text{End}\, \hG$ simply as $R(\X,c) = (R \X, c)$. The corresponding $R$-bracket on $\hG$ defined by the analog of \eqref{R-bracket} evaluates to
\begin{equation}  \label{R-bracket extension}
[(\X,a), (\Y,b)]_R \coloneqq \mbox{\small $\frac{1}{2}$} \left( [R (\X,a), (\Y,b)] + [(\X,a), R (\Y,b)] \right) = ([\X,\Y]_R, \omega_R(\X,\Y)),
\end{equation}
where $\omega_R(\X,\Y) \coloneqq \mbox{\small $\frac{1}{2}$} \left( \omega(R \X,\Y) + \omega(\X, R \Y) \right)$. Define the Kostant-Kirillov $R$-bracket on $\hGd_1$ as
\begin{equation} \label{Kostant-Kirillov R}
\mbox{\small $\frac{1}{2}$} \{ f, g \}_R(\hat{\L}) \coloneqq (\!( \hat{\L}, [ \hat{d}f, \hat{d}g ]_R )\!)_{\phi}.
\end{equation}
Here the factor of a half was introduced for convenience. Making use of \eqref{R-bracket extension} this can be evaluated explicitly to give
\begin{subequations} \label{Poisson operator R}
\begin{equation}  \label{Poisson operator R a}
\{ f, g \}_R (\hat{\L}) = - (\!( df, \mathcal{H}_R(\L) \cdot dg )\!)_{\phi},
\end{equation}
where the Poisson operator $\mathcal{H}_R(\L)$ is given by the skew-selfadjoint operator
\begin{equation} \label{Poisson operator R b}
\mathcal{H}_R(\L) \coloneqq ad \; \L \circ R + R^{\ast} \circ ad \; \L - (R + R^{\ast}) \partial_{\sigma}.
\end{equation}
\end{subequations}
The presence of the term in $\partial_{\sigma}$ stems once again from working with the central extension. Making use of the tensor notation in appendix \ref{app: Notations} to rewrite the bracket \eqref{Poisson operator R} yields
\begin{equation} \label{tensor R-bracket}
\{ f, g \}_R(\hat{\L}) = \int_{S^1} d\sigma \left( ( [R_{\1\2}, \L \otimes {\bf 1} ] - [R^{\ast}_{\1\2}, {\bf 1} \otimes \L ], df \otimes dg )_{\phi} + ( R_{\1\2} + R^{\ast}_{\1\2}, df \otimes \partial_{\sigma} dg )_{\phi} \right).
\end{equation}

\paragraph{The $r/s$-formalism.} The bracket \eqref{tensor R-bracket} is usually written for linear functions $f, g \in C^{\infty}(\hGd_1)$. So let $X, Y \in \Loop \gO$ and set $\X = X \cdot \delta_{\sigma_1}$, $\Y = Y \cdot \delta_{\sigma_2}$ where $\delta_{\sigma}$ denotes the Dirac $\delta$-function based at the point $\sigma \in S^1$. With these definitions we choose the linear functions
\begin{equation*}
f : \hat{\L} \mapsto (\!( \hat{\L}, (\X,0) )\!)_{\phi} = (\L(\sigma_1), X)_{\phi}, \qquad g : \hat{\L} \mapsto (\!( \hat{\L}, (\Y,0) )\!)_{\phi} = (\L(\sigma_2), Y)_{\phi}.
\end{equation*}
Now by abuse of notation the left hand side of \eqref{tensor R-bracket} is usually written replacing the functions $f, g$ by their values at $\hat{\L}$, or equivalently $\{ f, g \}_R(\hat{\L}) = (\{ \L \overset{\otimes}, \L \}_R, X \otimes Y)_{\phi}$ where it is understood that the first and second tensor factors depend on $\sigma_1$ and $\sigma_2$ respectively. By evaluating the right hand side of \eqref{tensor R-bracket}, using the standard notation $\L_{\1} = \L \otimes {\bf 1}$, $\L_{\2} = {\bf 1} \otimes \L$ for tensor products and the fact that $X, Y \in \Loop \gO$ are arbitrary, we obtain
\begin{equation} \label{r/s-bracket}
\{ \L_{\1} , \L_{\2} \}_R = [R_{\1\2}, \L_{\1}] \delta_{\sigma_1 \sigma_2} - [R^{\ast}_{\1\2}, \L_{\2} ] \delta_{\sigma_1 \sigma_2} + (R_{\1\2} + R^{\ast}_{\1\2}) \delta'_{\sigma_1 \sigma_2},
\end{equation}
where $\delta_{\sigma_1 \sigma_2} = \delta_{\sigma_2}(\sigma_1)$ and $\delta'_{\sigma_1 \sigma_2} = \delta'_{\sigma_2}(\sigma_1)$. Comparing this bracket to the standard $r/s$-algebra \cite{Maillet} we see that they are equivalent if we make the following identifications
\begin{equation} \label{r/s}
r \coloneqq \frac{1}{2}(R - R^{\ast}), \qquad s \coloneqq - \frac{1}{2}(R + R^{\ast}).
\end{equation}
In other words the $s$-matrix is nothing but (minus) the symmetric part of the $R$-matrix whereas the $r$-matrix corresponds to its skew-symmetric part \cite{Avan-et-al}. In particular the condition $s=0$ for ultralocality is equivalent to the skew-symmetry $R^{\ast} = - R$ of the $R$-matrix. The non-ultralocality of the supercoset $\sigma$-model can therefore be attributed to the twisted inner product \eqref{twisted ip}, which results in the non skew-symmetric condition \eqref{adj R}.
Substituting the kernels \eqref{proj kern resum twist 2} of $R$ and $R^{\ast}$ into the definitions
\eqref{r/s} we obtain the following kernels for the $r/s$-matrices,
\begin{subequations} \label{r-s Magro}
\begin{align}
\label{r Magro} r_{\1\2}(z_1, z_2) = \text{v.p.} &\frac{1}{z_2^4 - z_1^4} \left[ \sum_{j=0}^3 z_1^{4-j} z_2^j C^{(4-j \, j)}_{\1\2} \phi(z_1)^{-1} + \sum_{j=0}^3 z_1^j z_2^{4-j} C^{(j\, 4-j)}_{\1\2} \phi(z_2)^{-1} \right],\\
\label{s Magro} s_{\1\2}(z_1, z_2) = \, &\frac{1}{z_2^4 - z_1^4} \left[ \sum_{j=0}^3 z_1^{4-j} z_2^j C^{(4-j \, j)}_{\1\2} \phi(z_1)^{-1} - \sum_{j=0}^3 z_1^j z_2^{4-j} C^{(j\, 4-j)}_{\1\2} \phi(z_2)^{-1} \right].
\end{align}
\end{subequations}
Note that the expression \eqref{s Magro} is regular as $z_1 \rightarrow i^n z_2$ so the principal value is not needed for $s_{\1\2}$. If we neglect contact terms by omitting the principal value in $r_{\1\2}$ as well then these expressions are exactly the $r/s$-matrices of \cite{Magro}, up to an irrelevant overall factor of $4$.

\paragraph{Yang-Baxter equation.} The mCYBE \eqref{mCYBE tens 2} can also be rewritten in the equivalent form
\begin{equation} \label{mCYBE tens 3}
[R_{\tensor{12}}, (-R^{\ast})_{\tensor{13}}] + [(-R^{\ast})_{\tensor{12}}, (-R^{\ast})_{\tensor{23}}] + [(-R^{\ast})_{\tensor{13}}, (-R^{\ast})_{\tensor{23}}] = - \hat{\omega}_{\1\2\3}.
\end{equation}
Now using the definition \eqref{r/s} of the $r/s$-matrices we have $R = r - s$ and $-R^{\ast} = r + s$, and making use of \eqref{omega} for the right hand side we can rewrite \eqref{mCYBE tens 3} in full as follows,
\begin{multline*}
[r_{\tensor{13}}(z_1,z_3) + s_{\tensor{13}}(z_1,z_3), r_{\tensor{12}}(z_1,z_2) - s_{\tensor{12}}(z_1,z_2)] + [r_{\tensor{23}}(z_2,z_3) + s_{\tensor{23}}(z_2,z_3), r_{\tensor{12}}(z_1,z_2) + s_{\tensor{12}}(z_1,z_2)]\\
+ [r_{\tensor{23}}(z_2,z_3) + s_{\tensor{23}}(z_2,z_3), r_{\tensor{13}}(z_1,z_3) + s_{\tensor{13}}(z_1,z_3)] = - \omega_{\1\2\3} \, \phi(z_1)^{-2} \delta(z_1 - z_2) \delta(z_2 - z_3).
\end{multline*}
If we neglect the contact terms once more then the $\delta$-functions on the right hand side disappear, and we are left precisely with the so-called ``extended'' classical Yang-Baxter equation satisfied by the pair of matrices $r$ and $s$ \cite{Maillet}. We see that the correct form of this equation, including contact terms, is simply the mCYBE for a non skew-symmetric $R$-matrix.

\section{Lax matrix and coadjoint orbit} \label{sect: 6}

\paragraph{Coadjoint $R$-action.} In much the same way as the Kostant-Kirillov bracket \eqref{Kostant-Kirillov} generated flows \eqref{coadj orbit} along the coadjoint orbit of $\G$ in $\hGd_1$, the second Poisson bracket \eqref{Kostant-Kirillov R} will also generate flows along a different kind of coadjoint orbit in $\hGd_1$.

To see this, define the coadjoint action $ad_R^{\ast}$ of $\hG$ on $\hGd$ corresponding to the $R$-bracket \eqref{R-bracket extension} on $\hG$, in analogy with the definition \eqref{coadj action} of the coadjoint action $ad^{\ast}$ for the Lie bracket,
\begin{equation} \label{coadj action R}
(\!( ad_R^{\ast} (\M,c) \cdot (\X,a), (\Y, b) )\!)_{\phi} \coloneqq - (\!( (\X,a), [(\M,c),(\Y,b)]_R )\!)_{\phi}.
\end{equation}
As was the case for the coadjoint action \eqref{coadj action}, the right hand side is independent of $c \in \mathbb{C}$, which follows from \eqref{R-bracket extension}. The center of $\hG$ thus also acts trivially under $ad_R^{\ast}$, and hence the same expression \eqref{coadj action R} also defines the $ad_R^{\ast}$-action of $\G$ on $\hGd$, \textit{i.e.} the coadjoint action of $\G_R$ on $\hGd$.

Consider again the linear function given by $\hat{\X} \in \hG_1$. Its evolution under the flow $\partial_{\tau} \coloneqq \{ \cdot, h \}_R$ generated by any $h \in C^{\infty}(\hGd_1)$ reads $\partial_{\tau} (\!( \hat{\L}, \hat{\X} )\!)_{\phi} = \{ \hat{\X}, h \}_R(\hat{\L}) = 2 (\!( ad_R^{\ast} dh \cdot \hat{\L}, \hat{\X} )\!)_{\phi}$, which we can write simply as
\begin{equation} \label{coadjoint R-orbit}
\partial_{\tau} \hat{\L} = 2 \, ad_R^{\ast} dh \cdot \hat{\L}.
\end{equation}
In other words, the flow of any $h \in C^{\infty}(\hGd_1)$ with respect to the Poisson bracket \eqref{Poisson operator R} is tangent to the coadjoint orbit of $\G_R$ in $\hGd_1$.

\begin{lemma}
The $ad^{\ast}$- and $ad_R^{\ast}$-actions of $\G$ on $\hGd$ are related as
\begin{equation} \label{ad_R vs ad}
ad_R^{\ast} \X = \mbox{\small $\frac{1}{2}$} ad^{\ast} (R\, \X) + \mbox{\small $\frac{1}{2}$} R^{\ast} \circ ad^{\ast} \X, \qquad \X \in \G.
\end{equation}
\begin{proof}
This follows from the definitions \eqref{coadj action}, \eqref{R-bracket extension} and \eqref{coadj action R}.
\end{proof}
\end{lemma}

\paragraph{Zero-curvature equation.} The following important theorem provides a justification for having introduced the new Poisson bracket \eqref{Kostant-Kirillov R} based on the $R$-bracket.

\begin{theorem} \label{AKS thm}
The set $\mathcal{I}(\hGd_1)$ of Casimirs for \eqref{Kostant-Kirillov} are in involution with respect to \eqref{Kostant-Kirillov R} and the flow $\partial_{\tau} \coloneqq \{ \cdot, h \}_R$ generated by any $h \in \mathcal{I}(\hGd_1)$ satisfies the zero curvature equation
\begin{equation} \label{Lax}
\partial_{\tau} \L - \partial_{\sigma} \M = [\M, \L], \qquad \text{where} \quad \M = R (dh).
\end{equation}
\begin{proof}
Let $f,g \in \mathcal{I}(\hGd_1)$ be Casimirs of \eqref{Kostant-Kirillov}. Using \eqref{ad_R vs ad}, or directly from the definition \eqref{R-bracket extension} of the $R$-bracket one can show that
\begin{equation*}
\{ f, g \}_R(\hat{\L}) = (\!( ad^{\ast} dg \cdot \hat{\L}, R(\hat{d}f) )\!)_{\phi} - (\!( ad^{\ast} df \cdot \hat{\L}, R(\hat{d}g) )\!)_{\phi}.
\end{equation*}
But the property \eqref{coadj inv property} implies that $ad^{\ast} df \cdot \hat{\L} = ad^{\ast} dg \cdot \hat{\L} = 0$ and hence $\{ f, g \}_R = 0$.

Now making use of \eqref{ad_R vs ad} to rewrite equation \eqref{coadjoint R-orbit} yields
\begin{equation*}
\partial_{\tau} \hat{\L} = ad^{\ast} R(dh) \cdot \hat{\L} + R^{\ast} \circ ad^{\ast} dh \cdot \hat{\L}.
\end{equation*}
Since $h \in \mathcal{I}(\hGd_1)$ the last term vanishes by \eqref{coadj inv property} and using \eqref{coadjoint central} we obtain \eqref{Lax}.
\end{proof}
\end{theorem}

Notice that the flow $\partial_{\tau}$ not only lies in the coadjoint orbit of $\G_R$ in $\hGd_1$, from its definition \eqref{coadjoint R-orbit}, but it also lies in the coadjoint orbit of $\G$ in $\hGd_1$ from the fact that $\partial_{\tau} \hat{\L} = ad^{\ast} R(dh) \cdot \hat{\L}$.

Although theorem \ref{AKS thm} is quite general, it is also possible to generate zero curvature equations from functions which are not in $\mathcal{I}(\hGd_1)$. Indeed, this was the case for the zero curvature equation obtained in \cite{Ham} which is given by the following
\begin{lemma}
Let $P_- : \hat{\L} \mapsto - \mbox{\small $\frac{1}{4}$} (\!( \hat{\L}, \hat{\L} )\!)_{\phi}$ and $\partial_- \coloneqq \{ \cdot, P_- \}_R$. Although $P_- \not \in \mathcal{I}(\hGd_1)$ we still have,
\begin{equation*}
\partial_- \L - \partial_{\sigma} \L_- = [\L_-, \L], \qquad \text{where} \quad \L_- = \pi_- \L.
\end{equation*}
\begin{proof}
From the definition \eqref{Frechet der} we get $d P_- = - \mbox{\small $\frac{1}{2}$} \L$. Note also that due to the special form \eqref{Lax matrix} of the Lax matrix $\L$ and the expression \eqref{adj R} for the adjoint $R^{\ast}$ we have the relation $R^{\ast} \L = - \L$. Using all this we find $R^{\ast} \circ ad^{\ast} d P_- \cdot \hat{\L} = \mbox{\small $\frac{1}{2}$} (\partial_{\sigma} \L, 0)$. If $P_-$ were in $\mathcal{I}(\hGd_1)$ this would vanish. Nevertheless, the coadjoint $R$-action $ad_R^{\ast} d P_- \cdot \hat{\L} = ([\pi_- \L, \L] + \partial_{\sigma} (\pi_- \L), 0)$ still takes a zero curvature form.
\end{proof}
\end{lemma}

\paragraph{Lax matrix.} The upshot of theorem \ref{AKS thm} is the existence of integrable flows on $(\hGd_1, \{ \cdot, \cdot \}_R)$. Yet we are interested in the Hamiltonian dynamics of the supercoset $\sigma$-model which takes place on the phase-space $(\P, \{ \cdot, \cdot \})$ defined in \cite{Ham} (see section \ref{sect: 1}). The respective flows of the two systems can be related provided we have a Poisson map
\begin{equation} \label{Lax map}
\hat{\L} : (\P, \{\cdot, \cdot\}) \longrightarrow (\hGd_1, \{\cdot, \cdot\}_R),
\end{equation}
meaning that $\{ \hat{\L}^{\ast} f, \hat{\L}^{\ast} g \} = \hat{\L}^{\ast} \{ f, g \}_R$ for any $f,g \in C^{\infty}(\hGd_1)$. Indeed, the pullback by such a map of the flows on $\hGd_1$ then correspond exactly to the dynamical flows on $\P$. 
Note that once \eqref{Lax map} has been specified one can equally work at the level of $\hGd_1$.

The existence of such a Poisson map follows from the results of \cite{Ham, Magro} which we summarise in two propositions. Recall from section \ref{sect: 2} that the Lax matrix $\L$, defined by \eqref{Lax matrix}, takes values in the smooth dual $(\Loop \gO_-)^{\ast} \simeq \phi^{-1} (\Loop \gO_+)^{\perp}$. The following proposition shows that both this statement and the form \eqref{Lax matrix} of the Lax matrix $\L$ are invariant under the flows of theorem \ref{AKS thm}.

\begin{proposition} \label{prop: coadj orb}
The Lax matrix $\hat{\L} = (\L, 1)$ of the supercoset $\sigma$-model belongs to the coadjoint orbit $\hGd_{\sigma}$ of $\G_R$ inside $\hGd_- \simeq \phi^{-1} \G_+^{\perp} \oplus \{1\} \subset \hGd_1$ defined by $\L$ having the explicit form
\begin{equation} \label{Lax form}
\L = \phi(z)^{-1} \sum_{k=1}^{\infty} z^k (k\, A^{(k)} + B^{(k)}),
\end{equation}
for some $A, B \in C^{\infty}(S^1, \g)$.
\begin{proof}
All we need to show is that the form \eqref{Lax form} is invariant under the coadjoint $R$-action of $\G$. So letting $\X \in \G$ and $\L$ be as in \eqref{Lax form}, we need to show that $ad_R^{\ast} \X \cdot (\L,1) = (\L',0)$ with $\L'$ also of the form \eqref{Lax form}. Now using \eqref{ad_R vs ad} and \eqref{coadjoint central}, we have
\begin{equation*}
(\!( ad_R^{\ast} \X \cdot (\L,1), (\Y,b) )\!)_{\phi} = \mbox{\small $\frac{1}{2}$} (\!( [R \X, \L] + R^{\ast} [\X, \L] + (R + R^{\ast}) \partial_{\sigma} \X, \Y )\!)_{\phi}.
\end{equation*}
Let $\tilde{\L} = \sum_{k=1}^{\infty} z^{k-1} (k\, A^{(k)} + B^{(k)})$ so that $\L = \varphi^{-1} \tilde{\L}$. Then using \eqref{adj R} we can write
\begin{align*}
\mbox{\small $\frac{1}{2}$} ([R \X, \L] + R^{\ast} [\X, \L] + (R + R^{\ast}) \partial_{\sigma} \X) &= \varphi^{-1} \mbox{\small $\frac{1}{2}$} ([R \X, \tilde{\L}] - R [\X, \tilde{\L}] +(\varphi R - R \varphi) \partial_{\sigma} \X),\\
&= \varphi^{-1} ( \pi_- [\pi_+ \X, \tilde{\L}] - \pi_+ [ \pi_- \X, \tilde{\L}] + (\varphi \pi_+ - \pi_+ \varphi) \partial_{\sigma} \X),
\end{align*}
where in the second equality we have used the definition \eqref{R standard} of $R$. Now since $\tilde{\L} \in \G_+$, the first term disappears. Furthermore, since $\varphi = 16 z^{-1} \sum^{\infty}_{k = 1} k z^{4k}$ is a Taylor series in $z$, it follows that the last term only depends on the pole part $\X_- \coloneqq \pi_- \X$ of $\X$. Thus we are left with
\begin{align*}
\mbox{\small $\frac{1}{2}$} ([R \X, \L] + R^{\ast} [\X, \L] + (R + R^{\ast}) \partial_{\sigma} \X) = - \varphi^{-1} ( \pi_+ [ \X_-, \tilde{\L}] + \pi_+ (\varphi \partial_{\sigma} \X_-) ).
\end{align*}
Writing $\X_- = \sum_{i = 1}^{4 N} z^{-i} x_i$, where $x_i \in \g_{(-i)}$, one can show that both terms on the right hand side take the form $\sum_{k=1}^{\infty} z^{k-1} (k\, a^{(k)} + b^{(k)})$ for some $a,b \in C^{\infty}(S^1, \g)$, and the result follows. Letting
\begin{align*}
A'^{(k)} = \sum_{i=1}^{4 N} [x_i, A^{(k+i)}] + 4 \sum_{n = 1}^{N} x_{4n-(k)}, \quad B'^{(k)} = \sum_{i=1}^{4 N} [x_i, B^{(k+i)} - i A^{(k+i)}] + 4 \sum_{n = 1}^{N} (4n-(k)) x_{4n-(k)},
\end{align*}
we have $\L' = \phi(z)^{-1} \sum_{k=1}^{\infty} z^k (k\, A'^{(k)} + B'^{(k)})$.
\end{proof}
\end{proposition}

Recalling that the phase-space of the supercoset $\sigma$-model may be parametrised by the pair of fields $A_1, \nabla_1 \Pi_1 \in C^{\infty}(S^1, \g)$, the upshot of proposition \ref{prop: coadj orb} is that the map
\begin{equation*}
\hat{\L} : \P \longrightarrow \hGd_{\sigma},\quad (A_1, \nabla_1 \Pi_1) \longmapsto (\L, 1)
\end{equation*}
with $\L$ given by \eqref{Lax matrix}, is a bijection. Indeed, it is clearly injective and the surjectivity follows from proposition \ref{prop: coadj orb} because every point $\hat{\L} = (\L, 1)$ of the coadjoint orbit $\hGd_{\sigma}$, with $\L$ given by \eqref{Lax form}, is the image of $(\mbox{\small $\frac{1}{4}$} A, \mbox{\small $\frac{1}{8}$} B) \in \P$.
The result of \cite{Magro} can now simply be stated as
\begin{proposition} \label{prop: Lax}
The Lax matrix $\hat{\L} : (\P, \{ \cdot, \cdot \}) \rightarrow (\hGd_{\sigma}, \{ \cdot, \cdot \}_R)$ is a Poisson isomorphism.
\end{proposition}

In summary, by endowing the ``big'' space $\hGd_1$ with the Poisson structure $\{\cdot,\cdot\}_R$ constructed from a given choice of $R$-matrix \eqref{R standard}, we were able to ensure the existence of non-trivial integrable flows, according to theorem \ref{AKS thm}. Yet the phase-space $(\hGd_1, \{\cdot,\cdot\}_R)$, being parametrised by an infinite number of fields on $S^1$, is big enough to describe a very broad class of $2$-d integrable field theories. Indeed, different models with different field contents will correspond to different coadjoint orbits of $\G_R$ in $\hGd_1$. Proposition \ref{prop: Lax} states that the phase-space $(\P, \{ \cdot, \cdot \})$ of the supercoset $\sigma$-model can be identified with the coadjoint orbit $\hGd_{\sigma} \subset \hGd_1$ defined in proposition \ref{prop: coadj orb}.

\appendix

\section{Tensor notation} \label{app: Notations}

\paragraph{Tensor Casimir.} A useful object to consider is the so-called tensor Casimir $C \in \gcirc \otimes \gcirc$, defined as the image of $\text{id} \in \text{End } \gcirc \equiv \gcirc \otimes \gcirc^{\ast}$ under the isomorphism $\gcirc \otimes \gcirc^{\ast} \simeq \gcirc \otimes \gcirc$ induced by the inner product. In other words its defining property is
\begin{equation*}
\langle C_{\1\2}, x_{\2} \rangle_{\2} = x_{\1}, \quad \forall x \in \gcirc.
\end{equation*}
Moreover, we have the decomposition $C_{\1\2} = C^{(00)}_{\1\2} + C^{(13)}_{\1\2} + C^{(22)}_{\1\2} + C^{(31)}_{\1\2}$ where $C^{(i \, 4-i)}_{\1\2} \in \gcirc_{(i)} \otimes \gcirc_{(4-i)}$ since the inner product respects the grading.

\paragraph{Tensor kernels.} Given an operator $\O \in \text{End}\, \Loop \gO \simeq \Loop \gO \otimes (\Loop \gO)^{\ast}$ it will be useful to consider its kernel $\O_{\1\2} \in \Loop\gO \otimes \Loop\gO$ with respect to an inner product, say $(\cdot, \cdot)_{\phi}$, on $\Loop\gO$. It is defined by identifying $\Loop \gO \otimes (\Loop \gO)^{\ast}$ with $\Loop \gO \otimes \Loop \gO$ using the inner product through the relation
\begin{equation} \label{kernel}
(\O \,X, Y)_{\phi} \eqqcolon (\O_{\1\2}, Y \otimes X)_{\phi}, \qquad \forall \; X,Y \in \Loop \gO.
\end{equation}
Here we have extended the inner product to $\Loop \gO \otimes \Loop \gO$, first on pure elements in the obvious way $(A\otimes B, C \otimes D)_{\phi} = (A,C)_{\phi} (B,D)_{\phi}$ and then by linearity on others. We shall also use the notation $(A \otimes B, C)_{\phi \tensor{2}} = A (B,C)_{\phi}$ for a partial inner product over the second space only. For instance one can write the action of the operator in terms of its kernel as $(\O X)_{\1} = (\O_{\tensor{12}}, X_{\tensor{2}})_{\phi \tensor{2}}$.

It is easy to see from the definition \eqref{kernel} that the kernel $\O^{\ast}_{\1\2}$ of $\O^{\ast}$ is $P(\O_{\1\2}) = \O_{\2\1}$ where the linear map $P : \Loop \gO \otimes \Loop \gO \to \Loop \gO \otimes \Loop \gO$ is the permutation operator given by $P (A \otimes B) = B \otimes A$.

\section*{Acknowledgments}

I am grateful to M. Magro for carefully reviewing the first draft of this paper. I also thank F. Spill, C. A. S. Young and K. Zarembo for fruitful discussions. This work was supported by the ANR grant INT-AdS/CFT (ANR36ADSCSTZ).

\end{document}